\documentclass[aps, preprint, eqsecnum, secnumarabic, nofootinbib,
11pt]{revtex4}

\usepackage{graphicx}
\usepackage{epic}
\usepackage{eepic}
\usepackage{latexsym}
\usepackage{amsfonts}
\usepackage{amsmath}

\newcommand{\be}{\begin{equation}}
\newcommand{\ee}{\end{equation}}
\newcommand{\bea}{\begin{eqnarray}}
\newcommand{\eea}{\end{eqnarray}}
\newcommand{\del}{\partial}
\newcommand{\e}{\epsilon}

\newcommand{\vs}[1]{\vspace{#1 mm}}
\newcommand{\hs}[1]{\hspace{#1 mm}}

\def\a{\alpha}
\def\b{\beta}

\def\d{\delta}
\def\D{\Delta}
\def\e{\epsilon}

\def\f{\phi}

\def\m{\mu}
\def\n{\nu}

\def\th{\theta}

\def\O{\Omega}

\def\del{\partial}
\let\la=\label
\let\bm=\bibitem
\def\nn{\nonumber}

\begin{document}

\baselineskip=.50cm


\vs{3}

\title{A Note on Intersections of $S$-branes with 
$p$-branes}

\author{Nihat Sadik Deger}
\email[e-mail:]{ sadik.deger@boun.edu.tr}
\affiliation{Department of Mathematics, Bogazici University, \\
Bebek, 34342, Istanbul, Turkey \\} 
\affiliation{Feza Gursey Institute, \\
Cengelkoy, 34680, Istanbul, Turkey }

\date{\today}

\

\

\begin{abstract}
\baselineskip=.50cm \vs{3}
We first investigate intersections of an $S$-brane with a single $p$-brane and show that in addition to the already known solutions, it is possible to place the $S$-brane so that the radial part of the $p$-brane is not included in its worldvolume. This leads to a new set of solutions. Secondly, we consider intersections of an $S$-brane with a supersymmetric $Dp_1-Dp_2$ intersection and find the list of allowed solutions for both positions of the $S$-brane.
Among them there are $D1-D5-S1$ and $D1-D5-S5$ intersections which might be appropriate for studying time dependent $AdS/CFT$ correspondence.

\end{abstract}

\maketitle
\thispagestyle{empty}
\tableofcontents
\baselineskip=.55cm
\section{Introduction}
\setcounter{page}{1} 

Time dependent supergravity solutions recently attracted much attention due to their applications in cosmology. $S$-branes \cite{strominger} are particularly nice class of such solutions due to the fact that they have interpretation in string theory as $D$-branes with Dirichlet boundary conditions applied along the time-like direction. They may be useful in realizing the $dS/CFT$ correspondence \cite{ds} and are important for tachyon condensation \cite{tach1, tach2}. Therefore, it is desirable to increase the available supergravity $S$-brane solutions \cite{pope1, pope2, gutperle, myers, sol1, sol2, sol3, sol4, sol5, sol6, sol7, sol8, edel, sol9, sol10, sol11, sol12} which we aim to do in this paper.

In the past, intersecting $p$-brane solutions were studied substantially and it was understood that they 
can be constructed using the harmonic function rule \cite{har1,har2,har3,har4,har5, har6}. Later, it was found that $S$-brane intersections with \cite{sol1} and without \cite{sol4} supersymmetric analogues follow a similar procedure 
too. A natural next step was to consider intersections of $S$-branes with $p$-branes.
Assuming complete separation of variables in all physical fields and a factorized form for the spacetime metric, Edelstein and Mas were able to show that \cite{edel}  such intersections are obtainable from a similar set of rules. 
In the construction of \cite{edel} (which we call Class I) $S$-brane worldvolume was chosen to include the transverse space of the $p$-brane with the radial coordinate. In this paper, we will extend their work in two directions: first we will show that by not including the radial part of the $p$-brane in the worldvolume of the $Sq$-brane, one gets a new set of solutions (which we call Class II). However, in both cases $M5,M2$ and $D3$-brane solutions are allowed only with smearing which ruins their $AdS$ near horizon geometry. Nevertheless, there are $D5-S5$ (in Class I) and $D5-S1$ (in Class II) solutions without any smearing which motivates us to investigate the possibility of adding a $D1$-brane to these. Therefore, as a second generalization we study the intersections of an $S$-brane with a supersymmetric $Dp_1-Dp_2$ 
intersection and find the whole set of allowed solutions for both positions of the $S$-brane. We find that $D1-D5-S1$ and $D1-D5-S5$ solutions are allowed without any smearing for the $D5$-brane and therefore they have  $AdS_3 \times S^3$ near horizon limits as both radial and time coordinates approach to zero. We hope that these will be useful in understanding the $AdS/CFT$ duality \cite{mal1,mal2,mal3} in time dependent backgrounds.

The general structure of a $Dp_1-Dp_2-Sq$ intersection has the following form. A single $Sq$-brane solution with a flat transverse space is characterized with two functions, namely a hyperbolic function $H_q(t)$ and some  exponentials of time, whereas extremal $p$-brane solutions are described by a harmonic function, $H_p(r)$. Assuming a factorized form for the spacetime metric together with conditions for the dimensions of the smeared transverse spaces of the $p$-branes,
completely decouple radial and time derivatives in Einstein's equation. The same condition also solves the Einstein's equation corresponding to the only non-diagonal Ricci tensor component, $R_{0r}$, with an extra relation on integration constants. Dilaton is obtained simply from the superposition of the dilaton solutions of $p$ and $S$-branes and their gauge potentials get a multiplicative factor. Remaining equations are those of a $Dp_1-Dp_2$ intersection and a single $S$-brane. So, each section of the spacetime metric is multiplied by a factor of the form $H_{p_1}^{n_1}H_{p_2}^{n_2}H_q^{n_3}e^{n_4t}$ where the powers are fixed exactly as in the intersecting $Dp_1-Dp_2$ and a single $S$-brane solutions. The requirement that dimension of every group in the metric should be positive and the transverse dimensions of $p$ and $S$-branes should have certain lower bounds, restrict possible solutions. 

The organization of this paper is as follows. In the next section we begin our investigation of double intersections. 
After giving the general setup and a short review of intersecting $p_1-p_2$ and a single $S$-brane solutions,
we analyze in detail the condition for a solution to exist in both Class I and Class II. 
Then, in section 3 we do the same for the triple intersections and conclude in section 4 with some remarks and future directions. Technical details of the derivation of the solutions are presented in appendix A.

\section{Double Intersections}

Our action in the Einstein frame in $d$-dimensions is
\be
S=\int d^d x \sqrt{-g} \left( R- \frac{1}{2} \del_{\m}\f \del^{\m} \f - 
\sum_{A=1}^K \frac{1}{2(n_A)!}e^{a_A \f} F^2_{n_A} \right) \, , 
\ee
where Chern-Simons terms are omitted since they do not play any role in 
our solutions. To study 
intersections, we allow $K$ field strengths with degrees $n_A$ where $a_A$ 
is 
the dilaton coupling. 
This action is suitable for describing the bosonic parts of the low energy 
limits of type $IIA, 
IIB$
and 11-dimensional supergravities for some specific values of $a_A$ and 
$n_A$. Field equations are:
\bea
\la{eqn1}
&&R_{\m\n}=\frac{1}{2}\del_\m \f \del_\n \f +\sum_{A=1}^K 
\frac{1}{2(n_A)!}e^{a_A \f} \left( n_A F_{\m{\a_2}...{\a_{n_A}}} F_\n 
^{\, \a_2...\a_{n_A}} - \frac{(n_A -1)}{d-2}F^2_{n_A}g_{\m\n} \right) \,\, ,\\
\la{eqn2}
&&\del_\m \left( \sqrt{-g}e^{a_A \f} F^{\m \n_2 ...\n_{n_A}} \right)=0 \,\,\,\, , \\
\la{eqn3}
&&\frac{1}{\sqrt{-g}} \del_\m \left( \sqrt{-g}\del^\m \f 
\right)=\sum_{A=1}^K
\frac{a_A}{2(n_A)!}e^{a_A \f}F^2_{n_A} \,\,\,\, .
\eea
We also have the Bianchi identity $\del _{[\n}F_{\m_1...\m_{n_A}]}=0$.
For this action various $p$-brane (for a review see \cite{rev1, rev2, rev3}) and $S$-brane solutions have been 
constructed. The generic form of an intersecting, extremal $p_1-p_2$ brane solution is:
\bea
\la{solmet}
ds^2 &=& e^{2\a_1(r)} e^{2\a_2(r)}(-dt^2+ dx_1^2+...+dx_I^2) + e^{2\a_1(r)} e^{2\th_2(r)}(dy_1^2+...+dy_k^2) \nonumber \\
&+& e^{2\theta_1(r)} e^{2\a_2(r)} (dz_1^2+ ...+dz_b^2)  +
e^{2\theta_1(r)} e^{2\th_2(r)}(dr^2+ r^2d\O_m^2)  \,\,\,\, ,\\
\la{harmonic}
H_X &\equiv& \left[ 1 + \frac{\sqrt{\D_{X}}}{2(m-1)}\frac{Q_{X}}{r^{m-1}} \right] \,\,\,\, , \,\,\,\, \D_{X} \equiv a_{X}^2 + \frac{2(p_X+1)(d-p_X-3)}{d-2} \,\, \,\, , \,\,\,\, X=1,2 \,\, , \\
\la{radial}
e^{2\a_X(r)} &=& H_X^{-\frac{4(d-p_X-3)}{\D_{X} (d-2)}} \,\,\,\, , \,\,\,\, E_{X}^r = \frac{2}{\sqrt{\D_{X}}} H_X^{-1}
\,\,\,\,\,\,\,\, , \,\,\,\,\,\,\,\, X=1,2 \,\,\,\, , \\
\la{pscalar}
\f_{X} &=& - \frac{\e_{X} a_{X} (d-2)}{d-p_X-3} \a_X = \frac{\e_{X} a_{X} (d-2)}{p_X+1} \theta_X 
\,\,\,\, , \,\,\,\, X=1,2 \,\,\,\, , \,\,\,\, \f= \f_1+\f_2 \,\,\,\, ,
\eea
where $E_X^r$'s are the gauge potentials and $Q_{X}$'s are the charges. In supersymmetric solutions the dimension of the spatial intersection manifold is found to be   $I=(p_1+p_2)/2$ \,\, \cite{har1,har2,har3,har4,har5}.
For asymptotic flatness $m \geq 2$ is necessary. The electric and magnetic field strengths are given as
\be
F_{tx_1...x_{p_X}r}^X=\e_{x_1...x_{p_X}} \partial_r{E_X^r} \,\,\textrm{(Elec.)} \, , \, \hspace{0.2cm}
F^{x_{p_X+1}...\psi_1...\psi_m}_X = \frac{\e^{x_{p_X+1}...\psi_1...\psi_mr} e^{-a_X\f_X} \partial_r{E_X^r}}{\sqrt{-g}} \,\,\textrm{(Mag.)}
\ee
A single $p$-brane solution can be obtained from the above by setting the charge of the other brane to zero, which causes the corresponding harmonic function to become 1. However, this might produce some smearing.

On the other hand, 
a single $Sq$-brane solution with a flat, $l$ dimensional localized transverse space ($y_i$'s) and $k$ delocalized flat directions ($z_i$'s) is as follows 
\cite{gutperle, myers}:
\bea
\la{smetric}
ds^2&=&-e^{2A(t)} dt^2 + e^{2M(t)}(dx_1^2+..+dx_{q+1}^2) + e^{2N(t)}(dy_1^2+..+dy_l^2) 
+ e^{2R(t)}dz_idz_i
\\
\la{sscalar}
\f_q &=& -\frac{\e_q a_q(d-2)}{d-q-3}M(t) +c_1t +c_2 \,\,\,\,\, , \,\,\,\,\, \D_q \equiv a_q^2 + 
\frac{2(q+1)(d-q-3)}{d-2} \,\, , \\
\la{harmonic2}
e^{2M(t)} &=& \left(\frac{\sqrt{\D_q}}{2} \frac{Q_q}{c_3} e^{-\frac{\e_qa_q(c_1t+c_2)}{2}}
\cosh[c_3 (t-t_0)] \right)^{-\frac{4(d-q-3)}{\D_q(d-2)}} \equiv H_q^ {-\frac{4(d-q-3)}{\D_q(d-2)}} \,\,\,\, , \\  
\label{sson}
R &=& - \frac{(q+1)}{d-q-3}M(t) \,\,\, , \,\,\, A = \pm lt + R(t) \,\,\, , \,\,\,   N= \pm t + R(t)    \,\, , \\
\la{ssonson}	
E_q^t &=& \frac{4c_3 e^{c_3(t-t_0)}}{Q_q\D_q\cosh[c_3(t-t_0)]} \,\, ,
\eea
where $c_1, c_2$ and $c_3$  are constants that satisfy
\be
(q+1)c_1^2+\frac{2(d-2)}{d-q-3}c_3^2 - \frac{(d-2)l(l-1)\D_q}{d-q-3}=0 \, ,
\la{sconstraint}
\ee
from which it follows that the dimension of the localized transverse space (that is $l$) should be greater than 1, that forces $q\leq 6$. (This restriction can be circumvented as explained in \cite{myers}.)
The constant $Q_q$ is the charge of the $Sq$-brane. The limit $c_3=0$ is achieved by setting $Q_q=0$ and
$c_3/Q_q=1$ in the solution given above (\ref{smetric})-(\ref{ssonson}). In this case $H_q$ takes a simpler form.
The field strengths are  
\be
F_{tx_1...x_{q+1}}=\e_{x_1...x_{q+1}} \partial_t{E_q^t}  \,\,\,\,\textrm{(Electric)} \, , 
\hspace{0.2cm} F^{y_1...y_lz_1...z_k} = \frac{\e^{y_1...y_lz_1...z_kt} e^{-a_q\f_q} \partial_t{E_q^t}}{\sqrt{-g}} \,\,\,\,\textrm{(Magnetic)}
\ee
Value of the dilaton couplings '$a_p$' and '$a_q$' for a $Dp$ and $Sq$-brane 
in type $IIA$ and $IIB$ 
supergravities are given by:
\be
\begin{cases} \e_p a_p=\frac{3-p}{2} \, , \hs{10} \e_q a_q= \frac{3-q}{2} \hspace{1cm}
\textrm{$RR$-branes} \cr
\e_p a_p=\frac{p-3}{2} \, , \hs{10} \e_q a_q= \frac{q-3}{2} \hspace{1cm}
\textrm{$NS$-branes}
\end{cases}
\ee
where $\e_p=\e_q=1$ for electric branes ($p=0,2$, $q=0,2$ in type $IIA$ 
and 
$p=1$, $q=1$ in type 
$IIB$) and $\e_p=\e_q=-1$ for magnetic branes ($p=4,6$ , $q=4,6$ in type 
$IIA$ 
and $p=5$, $q=5$ in type $IIB$). In 11-dimensional supergravity there is no dilaton 
and \, $a_q=a_p=0$. With these values, in 11-dimensions and in type $IIB$ and $IIA$ theories we have $\D_p=\D_q=4$. 

To find solutions for intersections of an $Sq$-brane with 
a single $Dp$-brane our metric ansatz is:
\bea
ds^2=- e^{2A(t)} e^{2\a (r)} dt^2 &+& e^{2B(t)} e^{2\a 
(r)}(dx_1^2 +... + dx_I^2) + e^{2C(t)} e^{2\a (r)} (dy_1^2 + ... +dy_k^2) \nn \\ 
&+& e^{2E(t)} e^{2\theta (r)} (dz_1^2 + ... +dz_b^2) 
+ e^{2D(t)} e^{2\theta (r)} (dr^2+ r^2d\O _m ^2)
\label{metric}
\eea
The coordinates $(t, x_1, .., x_I, y_1,...., y_k)$ build up the
worldvolume of the $Dp$-brane and therefore $p=k+I$. The  directions $(z_1, ..., z_b)$  are delocalized for the $Dp$-brane which has $(m+1)$ dimensional transverse space. It follows that $d=p+b+m+2$. Here $(x_1,...,x_I)$ is the intersection manifold with the $Sq$-brane. No other division in the metric is necessary. As shown in appendix A, the radial metric functions are those of a single $p$-brane, namely 
$e^{2\a}=H_p^{-\frac{d-p-3}{d-2}}$ and $e^{2\th}= H_p^{\frac{p+1}{d-2}}$ where the harmonic function $H_p$ is given in (\ref{harmonic}). For the scalar field we have
\be
\f =   \f_p + \f_q = - \frac{\e_p a_p (d-2)}{d-p-3} \a(r)  - \frac{\e_q a_q(d-2)}{d-q-3}B(t) +c_1t +c_2 \,\, .
\la{dilaton}
\ee
This is quite a natural result since we look for localized 
solutions in the sense that, when a brane is removed by setting its charge to zero we still have a solution for the other. We assume separation of $t$ and $r$ variables also for the gauge potentials and get (\ref{gauge1})-(\ref{gauge2}).

In order to have decoupling between time and radial derivatives in Einstein's equations, one finds that some restrictions have to be imposed (see the appendix (\ref{sep2})-(\ref{sep3}) ). It turns out that these are exactly what one needs in order to be able solve the Einstein's field equation for the non-diagonal component of the Ricci tensor. In the orthonormal frame 
for the above metric (\ref{metric}) it is given as
\be
e^{A+D} e^{\a + \theta} R_{0r}= (I+k+m) \a ' \dot{D}  
+ b  [\a'\dot{E}+ \th'(\dot{D}-\dot{E})] 
\,\, . \\
\la{nondia}
\ee
Now inserting our dilaton solution (\ref{dilaton}) in Einstein's equation (\ref{eqn1}) for the $R_{0r}$ component, gives restrictions for the smearing dimension $b$ and consequently for the intersection dimension $I$. Also, we get an additional condition on integration constants.

There are two ways to place the remaining directions of the $Sq$-brane: the 
transverse space of the $p$-brane (i.e. $e^{2D}$ part of the metric) is either 
orthogonal to the $Sq$-brane or it is
included in its worldvolume. The first case was considered in \cite{edel}. Hence, we will focus on the second case, but for completeness we will summarize
the first type of solutions as well. Details of the derivation of the solutions can be found in appendix A. Let us now analyze  these two cases separately:

\subsection{Class I}

In this case, the radial part of the $p$-brane solution, $\Sigma_{m+1}$, is contained in the worldvolume of the $Sq$-brane which is constructed from $(x_1, ..., x_I, \Sigma_{m+1})$ and hence $q=I+m$ and $d=q+b+k+2$. Here we have $e^{2B}=e^{2D}=H_q^{-\frac{d-q-3}{d-2}}$ and $e^{2C}=e^{\pm2t}H_q^{\frac{q+1}{d-2}}$ where $H_q$ is given in (\ref{harmonic2}). If $e^{2E(t)}$ part of the metric is delocalized for the $Sq$-brane we have 
$e^{2E} = H_q^{\frac{q+1}{d-2}}$ and $e^{2A} = e^{\pm 2kt} H_q^{\frac{q+1}{d-2}}$, otherwise $e^{2E}=e^{\pm2t}H_q^{\frac{q+1}{d-2}}$ and $e^{2A} = e^{\pm 2(k+b)t} H_q^{\frac{q+1}{d-2}}$.
Using (\ref{dilaton}) in (\ref{nondia}) we find
\be
b = \frac{(d-p-3)(d-q-3)}{d-2} - \frac{\e_pa_p\e_qa_q}{2}\, , 
\label{double2}
\ee
from which we deduce the dimension of the intersection manifold
\be
I = \frac{(p+1)(q+1)}{d-2} - \frac{\e_pa_p\e_qa_q}{2} -2 \, ,
\ee
which is already obtained in \cite{edel}. If the constant $c_1 \neq 0$, then directions multiplied by $e^{2E(t)}$ are not smeared for the $Sq$-brane and we get the constraint
\be
\e_pa_pc_1=\mp 2b \,\, .
\label{constant2}
\ee
If $c_1=0$, like in $d=11$, then we don't have this condition and directions multiplied by $e^{2E(t)}$ have to be smeared for the $Sq$-brane.
Since $I \geq 0$, in 11-dimensions we should have $(p+1)(q+1) \geq 18$ which prohibits $M2-SM2$ intersection.
Conditions $k\geq 2$ and $m \geq 2$ are satisfied in all three intersections given below where there is always smearing (i.e. $b \neq 0$) and \, $9(m-1)= (q+1)(8-p)$.

In 10-dimensions, the above expressions take the form $b= (10-p-q)/2$ and $I=(p+q-6)/2$. Thus, we have the restriction $6 \leq (p+q) \leq 10$. In order to have a harmonic function for the $Dp$-brane we need $m \geq 2$ which happens when $(p-q) \leq 2$. (Intersections are possible without this restriction, but then $H_p$ is no longer a harmonic function.) 
The dimension of the transverse space of the $Sq$-brane should satisfy $(k+b) \geq 2$  giving $q\leq 6$ which is always true. Note that there will be no smearing for the $Dp$-brane when 
$(p+q)=10$ which is fulfilled only for $D4-S6$, 
$D6-S4$ and $D5-S5$ intersections. Repeating the same analysis for $NS$-branes we obtain
\bea
&&(0|M5,SM2|2) \nonumber \\
&&(0|M2,SM5|2) \nonumber \\
&&(2|M5,SM5|1) \nonumber \\
&&(\frac{p+q}{2}-3|Dp,Sq|5-\frac{p+q}{2})  \hspace{1cm} 6 \leq (p+q)\leq 10 \,\,\, , \,\,\, 2 \geq (p-q) \nonumber \\
&&(p-2|Dp,SNS5|1) \nonumber \\
&&(q-2|NS5,Sq|1) \nonumber \\
&&(2|NS5,SNS5|0) \nonumber \\
&&(0|NS1,SNS5|2) 
\la{list4}
\eea
following the notation $(I|Dp,Sq|b)$ introduced in \cite{edel}. All the solutions given in this subsection 
have already appeared in \cite{edel} except the first two and the last one on the list (\ref{list4}).

\subsection{Class II}
\label{sect1}

In this case the transverse part of the $p$-brane (i.e. $e^{2D}$ part of the metric) is not included in the worldvolume of the $Sq$-brane. Note that the $b$-dimensional space is smeared for the $p$-brane but it is part of the $Sq$-brane worldvolume which consists of $(x_1, ..., x_I, z_1,...,z_b)$. We have $e^{2B}=e^{2E}=H_q^{-\frac{d-q-3}{d-2}}$ and $e^{2C}=e^{\pm2t}H_q^{\frac{q+1}{d-2}}$ where $H_q$ is given in (\ref{harmonic2}). If $e^{2D(t)}$ part of the metric is delocalized for the $Sq$-brane we have 
$e^{2D} = H_q^{\frac{q+1}{d-2}}$ and $e^{2A} = e^{\pm 2kt} H_q^{\frac{q+1}{d-2}}$, otherwise $e^{2D}=e^{\pm2t}H_q^{\frac{q+1}{d-2}}$ and $e^{2A} = e^{\pm 2(k+m+1)t} H_q^{\frac{q+1}{d-2}}$.
Here $q=I+b-1$ and $d=q+k+m+3$. Using (\ref{dilaton}) and (\ref{nondia}) in (\ref{eqn1}) we find
\be
b = \frac{(d-p-3)(q+1)}{d-2}  +\frac{\e_pa_p\e_qa_q}{2} \,\, ,
\label{double1}
\ee
from which we deduce the dimension of the intersection manifold
\be
I = \frac{(p+1)(q+1)}{d-2} - \frac{\e_pa_p\e_qa_q}{2}
\,\, .
\ee
If the constant $c_1 \neq 0$, then directions multiplied by $e^{2D(t)}$ are not smeared for the $Sq$-brane
and we get the constraint
\be
\e_pa_p c_1=\mp 2(d-p-b-3) = \mp 2(m-1) \,\, .
\label{constant}
\ee
If $c_1=0$, such as in $d=11$,  then we don't have this condition and directions multiplied by $e^{2D(t)}$ are smeared for the $Sq$-brane. Sometimes, such as in $D2-S4$ intersection, one is obliged to do this in order to be able to solve the constraint (\ref{sconstraint}).

In 11-dimensions $b=(8-p)(q+1)/9$ and $I=(p+1)(q+1)/9$. To have a valid $Sq$-brane solution we need to have $k\geq 2$ which happens only in $M5-SM2$ intersection where $m=3$.

In 10-dimensions $b=(4+q-p)/2$ and $I=(p+q-2)/2$. 
It follows that a solution exists only when $(p-q) \leq 4$ and $p \geq |q-2|$  where we also used $p \geq I$. In order to have a harmonic function for the $Dp$-brane we impose $m\geq 2$ which gives $8\geq (p+q)$. For the $Sq$-brane we need
to have $(k+m) \geq 1$ which happens for $q \leq 6$. The only exception is the $D3$-brane where only $q=1$ is allowed due to the fact that we must have $k\geq 2$.
There will be no smearing (i.e. $b=0$) for the $Dp$-brane when $(p-q)=4$ which is possible only for $D4-S0$, $D6-S2$ and $D5-S1$ intersections. In these, the $Sq$-brane is contained completely inside the $Dp$-brane. 
Repeating the same analysis for $NS$-branes we finally obtain 
\bea
&&(2|M5,SM2|1) \nonumber \\
&&(\frac{q+p}{2}-1|Dp,Sq|2+\frac{q-p}{2})  \hspace{1cm} |q-2| \leq p\leq (4+q) \,\,\, , \,\,\, p \neq 3 \nonumber \\
&&(1|D3,S1|1) \nonumber \\
&&(p|Dp,SNS5|6-p) \hspace{1cm} p \neq 3 \nonumber\\
&&(q|NS5,Sq|1) \hspace{1.4cm}  \nonumber \\
&&(1|Dp,SNS1|1) \hspace{1.4cm}  \,\,\,\, 1 \leq p \leq 5 \nonumber \\
&&(1|NS1,Sq|q) \hspace{1.4cm}   \nonumber \\
&&(0|NS1,SNS1|2) 
\la{list1}
\eea
in the notation $(I|Dp,Sq|b)$.

\section{Triple Intersections}

The above analysis showed that intersections involving $M2$, $M5$ or $D3$ branes are possible only with smearing. Therefore, these are not appropriate for studying $AdS$ near horizon geometries. However, there exists $D5-S1$ and $D5-S5$ solutions without any delocalized directions for the $D5$-brane. Since the near horizon limit of $D5-D1$ intersection is $AdS_3 \times S^3$, it is desirable to find out whether it is possible to add an $S$-brane to such an intersection. Therefore, in this section we investigate intersections of an $Sq$-brane with a $Dp_1-Dp_2$ intersection. In an $Dp_1-Dp_2$ intersection, worldvolume of each $Dp$-brane is divided into two parts as intersecting and non-intersecting directions. Each one of these should be 
divided into two more parts when an $Sq$-brane is added. Therefore, worldvolume of each $Dp$-brane has 4 pieces. Our metric ansatz is
\bea
ds^2= &-& e^{2A(t)} e^{2\a (r)} dt^2 + e^{2B(t)} e^{2\a 
(r)}(dx_1^2 +... + dx_n^2) + e^{2C(t)} e^{2\a (r)} (dy_1^2 + ... +dy_k^2) \nn \\ 
&+& e^{2E(t)} e^{2\b (r)} (dw_1^2 + ... +dw_j^2) + e^{2F(t)} e^{2\b (r)} (dz_1^2 + ...+dz_i^2)
+ e^{2G(t)}e^{2\psi (r)}(du_1^2+...+du_a^2) \nn \\ &+& e^{2U(t)}e^{2\psi (r)}(dv_1^2+...+dv_b^2)
+ e^{2D(t)} e^{2\theta (r)} (dr^2+ r^2d\O _m ^2) \,\, .
\label{metric2}
\eea
Here the $Dp_1$-brane worldvolume coordinates are $(t, u_1,...,u_a, v_1,...,v_b, x_1,...x_n, y_1,...y_k)$
and $Dp_2$-brane is located at $(t, x_1,...,x_n, y_1,...,y_k,z_1,...,z_i, w_1,...,w_j)$.
Thus, we have $p_1=a+b+n+k$ and $p_2= i+j+n+k$ in addition to $(n+k+i+j+a+b+m)=d-2$. Note that there are $(i+j)$ smeared directions for $Dp_1$ and $(a+b)$ smeared directions for $Dp_2$ and we don't allow an overall smearing for them. The radial metric functions are \, $\a=\a_1+\a_2 \, , \, \b = \a_1+\th_2 \, , \, \psi=\th_2 + \a_2 \, , \, \th= \th_1+\th_2$ \, where $e^{2\a_X}=H_X^{-\frac{d-p_i-3}{d-2}}$ and $e^{2\th_X}= H_X^{\frac{p_i+1}{d-2}}$ \, $X=1,2$ \, where $H_X$ functions are given in (\ref{harmonic}).
For the time dependent metric functions we have the following rule; the worldvolume directions of the $Sq$-brane are multiplied by $e^{2B}=e^{2E}=e^{2U}=H_q^{-\frac{d-q-3}{d-2}}$ and localized transverse directions of the $Sq$-brane are multiplied by $e^{2C}= H_q^ {\frac{q+1}{d-2}}$ where $H_q$ is given in (\ref{harmonic2}). Delocalized transverse directions have the factor $ e^{\pm2t} H_q^ {\frac{q+1}{d-2}}$ and we have $e^{2A}=e^{\pm2lt} H_q^ {\frac{q+1}{d-2}}$ where $l$ is the total number of localized transverse directions for the $Sq$-brane. The non-diagonal part of the Ricci tensor is
\bea
e^{A+D} e^{\a + \theta}R_{0r} &=& (n+k+m)\a'\dot{D} + j[\a'\dot{E} + \b'(\dot{D}-\dot{E})] + i[\a'\dot{F} + \b'(\dot{D}-\dot{F})] \nn \\
&+& a[\a'\dot{G} + \psi'(\dot{D}-\dot{G})] + b[\a'\dot{U} + \psi'(\dot{D}-\dot{U})] \,\, .
\label{nondia2}
\eea

For the dilaton we have the same solution as above (\ref{dilaton}). The gauge potentials are given in (\ref{gauge1})-(\ref{gauge2}). Once again there are two options for the worldvolume of the $Sq$-brane; it either includes the overall transverse space of the $Dp_1-Dp_2$ intersection or not:

\subsection{Class I}

We assume that $e^{2D}$ part of the metric, $\Sigma_{m+1}$, is contained in the worldvolume of the $Sq$-brane. Therefore, it is oriented along $(x_1,...,x_n, w_1,...,w_j, v_1,...,v_b, \Sigma_{m+1} )$ and $q=n+j+b+m$.
We have $D(t)=U(t)=E(t)=B(t)$. From (\ref{nondia2}) and (\ref{dilaton}) one finds that
\bea
\label{triple21}
a &=& \frac{(d-p_2-3)(d-q-3)}{d-2} - \frac{\e_{p_2}a_{p_2}\e_qa_q}{2}\,\, , \\
\label{triple22}
i &=& \frac{(d-p_1-3)(d-q-3)}{d-2} - \frac{\e_{p_1}a_{p_1}\e_qa_q}{2}\,\, .
\eea
When there is a single $Dp$-brane we get only one of the above equations and the remaining one is equivalent to (\ref{double2}) after appropriate relabelling, as it should be. The directions multiplied by $e^{2F(t)}$ and $e^{2G(t)}$ are either smeared for the $Sq$-brane or transverse. If both of them are smeared then we have $c_1=0$, otherwise we have the constraints 
\be
(3-p_1)c_{p_1} = \mp 4i \, , \hspace{1cm} (3-p_2)c_{p_2} = \mp 4a \, ,
\la{constant4}
\ee
and $c_1= c_{p_1}+c_{p_2} $ which is the generalization of (\ref{constant2}).

In 11-dimensions one finds $k-1=(8-q)(p_1+p_2-7)/9$. From the condition $k \geq 2$ \, we see that only $M5-M5-SMq$ \, intersection is allowed with both $q=2$ and $q=5$
where $n=3-k, \, j=b=k-1, \, a=i=(8-q)/2, \, m=2$.

In 10-dimensions we find $k+2=(p_1+p_2)/2$. In supersymmetric intersections we have \cite{har1,har2,har3,har4,har5} $k+n+2=(p_1+p_2)/2$ which implies that $n=0$, i.e. there is no common intersection of all three branes. Other unknowns are found to be
$a=(10-p_2-q)/2, \, i=(10-p_1-q)/2, b=2-i, \, j=2-a, \, m=(12-p_1-p_2)/2$. In a given $Dp_1-Dp_2$ intersection all $q$'s are allowed provided that $(6-p_{min}) \leq q \leq (10-p_{max})$. Additionally we need to have $8 \geq (p_1+p_2)$ for $m \geq 2$. The requirement $(k+i+a) \geq 2$ gives the trivial condition $q \leq 6$. This section can be summarized as
\bea
\la{list3}
&& (\frac{q+1}{3}|M5,M5,SMq|2) \hspace{1cm} q=2\,\, , \,\,  q=5  \\
&& (0|Dp_1,Dp_2,Sq|6- \frac{p_1+p_2}{2}) \hspace{0.5cm} 
(6-p_{min}) \leq q \leq (10-p_{max}) \,\, , \hspace{0.3cm}  4  \leq (p_1+p_2) \leq 8 \nonumber
\eea
in the notation $(n|Dp_1,Dp_2,Sq|m)$. Observe that 
we can add an $S5$-brane to the  $D1-D5$ intersection without any smearing for the $D5$-brane. Spatial directions are splitted as 1+2+2+4. Since $k=1$, directions multiplied by $e^{2F}$ should be localized for the $S5$-brane so that its overall transverse space is three dimensional. From the constraint (\ref{sconstraint}) we find $c_3=0$ which  forces to set $Q_q=0$ and $c_3/Q_q=1$ in the solution given in (\ref{smetric})-(\ref{ssonson}).

\subsection{Class II}

Now let us assume that $e^{2D}$ part of the metric is outside of the $Sq$-brane. Then, the worldvolume of the $Sq$-brane is $(x_1,...,x_n, w_1,...,w_j, v_1,...,v_b)$ and so $q=n+j+b-1$. From (\ref{nondia2}) and (\ref{dilaton}) one finds that
\bea
\label{triple11}
b &=& \frac{(d-p_2-3)(q+1)}{d-2}  +\frac{\e_{p_2}a_{p_2}\e_qa_q}{2} \,\, , \\
\label{triple12}
j &=& \frac{(d-p_1-3)(q+1)}{d-2}  +\frac{\e_{p_1}a_{p_1}\e_qa_q}{2} \,\, .
\eea
When there is only one $Dp$-brane such as $Dp_2$ (here we have $a=k=0$) we don't have (\ref{triple12}) and equation (\ref{triple11}) with $p_2=p$ is nothing but our previous result for a single $Dp$-brane (\ref{double1}), as it should be. The directions multiplied by $e^{2D(t)}, e^{2F(t)}$ and $e^{G(t)}$ are either smeared for the $Sq$-brane or localized transverse. If all of them are smeared then we have $c_1=0$, otherwise we have the constraints 
\be
(3-p_1)c_{p_1}=\mp 4(d-p_1-3-j) \, , \hspace{1cm} (3-p_2)c_{p_2}=\mp 4(d-p_2-3-b) \,\, ,
\la{constant3}
\ee
and  $c_1= c_{p_1} + c_{p_2}$ \, , which reduces to (\ref{constant}) when there is only a single $Dp$-brane. Note that for $Dp_1$ (respectively $Dp_2$) there are $j$ (respectively $b$) smeared directions intersecting with the $Sq$-brane. 

In 11-dimensions from the above we obtain $9n=(q+1)(p_1+p_2-7)$. Hence, we should have $(p_1+p_2) \geq 7$ which prohibits $M2-M2$ possibility. The restriction 
$k\geq 2$ \, rules all but one intersection namely $M5-M5$ with $SM2$ where we have
 $a=b=i=j=n=1$ and $m=k=2$. 

In 10-dimensions from (\ref{triple11}) and (\ref{triple12}) we find $n+3=(p_1+p_2)/2$. Therefore, $p_1+p_2 \geq 6$ which rules out some supersymmetric intersections. Let us now focus on supersymmetric intersections where we have \cite{har1,har2,har3,har4,har5} \, $k+n+2=(p_1+p_2)/2$ which implies that $k=1$. Hence, there is no solution with $D3$-branes. Other unknowns are found to be $a=(p_1-q)/2, \, i=(p_2 -q)/2 , 
\, j=2-a, \, b=2-i, \, m=6-(p_1+p_2)/2$. We can combine the inequalities that arise from these as 
$(p_{max}-4) \leq q \leq p_{min}$ where $p_{min}$ (respectively $p_{max}$) is the minimum (respectively maximum) of $p_1$ and $p_2$. (From these it also follows that $p_{min} \geq |q-2|$ as in section \ref{sect1}.)  The requirement $m \geq 2$ implies 
$8 \geq (p_1+p_2)$ and the transverse space of the $Sq$-brane automatically has dimension greater than 1. Therefore, once $p_1$ and $p_2$ with $ 8 \geq (p_1+p_2) \geq 6$ are chosen  
all $q$'s satisfying the above inequality are allowed (of course, $q$ has to be even in type $IIA$ and odd in type $IIB$).  In summary, the allowed intersections are
\bea
\la{list2}
&& (1|M5,M5,SM2|2)  \\
&& (\frac{p_1+p_2}{2}-3|Dp_1,Dp_2,Sq|6- \frac{p_1+p_2}{2}) \hspace{0.3cm} 6\leq (p_1+p_2) \leq 8 \,\, , \hspace{0.1cm} (p_{max}-4) \leq q \leq p_{min} \,\, , p_X \neq 3  \nonumber
\eea
in the notation $(n|Dp_1,Dp_2,Sq|m)$ where $X=1,2$. Note that, this shows that we can add an $S1$-brane to the  $D1-D5$ intersection without any smearing for the $D5$-brane. Spatial directions are splitted as 1+2+2+4.
When $e^{2D}$ part is chosen to be delocalized for the $S1$-brane we find $c_3=0$ from (\ref{sconstraint}) which can be handled only by setting its charge to zero. However, if $e^{2D}$
part of the metric is localized transverse space for the $S1$-brane then it has a nonzero charge.

\section{Conclusions}

In this paper we studied intersections of $S$-branes with $p$-branes. The set of new solutions that we found are listed in (\ref{list1}), (\ref{list3}) and (\ref{list2}) which are the main results of this work.
Let us emphasize that some intersections appear on both types but with different amounts of smearings. For example, $D6-S4$ intersection has $b=3, I=4$ in Class II and $b=0, I=2$ in Class I. 

Some of the intersections considered here can directly be obtained from $d=11$. For instance, we can take the standard $SM5$ solution and replace 4-flat directions in its worldvolume with the Taub-NUT space which is allowed by the Einstein's equations since it is a Ricci-flat space. Then, using a particular coordinate of this space for dimensional reduction we find $D6-S4$ solution  which is given in section 2.1. If instead we use the Taub-NUT space in the transverse part of the $SM5$ and $SM2$, then we get $D6-SNS5$ and $D6-S2$ intersections respectively, that are given in section 2.2. In these, there is no smearing 
for the $D6$-brane.

It is straightforward to generalize our method to include additional $S$-branes. However, from our tables  (\ref{list4}), (\ref{list1}), (\ref{list3}) and (\ref{list2}) it is easy to see that adding more $S$-branes to $D1-D5-S1$ and $D1-D5-S5$ is not possible. Another easy modification would be to consider hyperbolic and spherical transverse spaces for the $Sq$-brane which might be useful in cosmological applications of these solutions.

Among our solutions perhaps the most interesting ones are $D1-D5-S5$ and $D1-D5-S1$ intersections due to the fact that they have $AdS_3 \times  S^3$ limit as $r \rightarrow 0$ and $t \rightarrow 0$. In the first case the $S5$-brane turns out to be neutral, whereas in the latter solution both charged and chargeless $S1$-branes are allowed.
These intersections might be used in studying $AdS/CFT$ \cite{mal1, mal2, mal3} duality in time dependent backgrounds
but certainly more work is required in order to clarify their geometric structure. It would be very interesting to try to repeat the analysis of \cite{cvetic} which was done for a cosmological $AdS_5$ space, such as calculation of conformal anomaly. Setups similar to ours in the context of type $II$ supergravities have already been considered recently \cite{chu, nay, oh, das}. Unlike these, in our solutions presence of the $S$-brane destroys the supersymmetry. Therefore, techniques employed in \cite{horo} might be more relevant. An $Sq$-brane gives rise to 
an accelerating cosmology in $(q+1)$ dimensions after 
compactification \cite{inf1,inf2,inf3,inf4,neu1}. In this regard, $D6-S2$ and $D6-D2-S2$ intersections found in sections 2.2 and 3.2 seem especially attractive. Our solutions provide a fascinating opportunity 
to study inflation from the $AdS/CFT$ point of view \cite{alb,inf5,neu2}, which we hope to explore soon.

\begin{acknowledgments}
It is a pleasure to thank A. Kaya for
discussions. I would like to thank the Abdus Salam ICTP for
hospitality where some part of this paper was written. This work is partially
supported by Turkish Academy of Sciences via The Young Scientist Award Program (T{\"U}BA-GEB{\.I}P).
\end{acknowledgments}

\appendix

\section{Derivation of The Solutions}

We consider a metric of the following form
\be
ds^2=-e^{2A(t)} e^{2\a(r)} dt^2\,+\,\sum_k\,e^{2C_k(t)} e^{2\b_k(r)}\,ds_k^2\,+\,e^{2D(t)} e^{2\theta(r)}\,(dr^2 + r^2d\Omega_m^2)
\ee
where $ds_k^2$ (for $k=1,2...$) is the metric on the $d_k$ dimensional flat
space. With respect to the the
orthonormal frame \, $\tilde{E}^{0}=e^{A(t)} e^{\a(r)}dt$, $\tilde{E}^{x_k}=e^{C_k(t)} e^{\b_k(r)}dy^{x_k}$,
$\tilde{E}^{r}=e^{D(t)}e^{\theta(r)}dr$, \, and \, $\tilde{E}^{\psi}= r e^{D(t)}e^{\theta(r)} e^{\psi}$ where $e^{\psi}$ is an orthonormal frame on $\Omega_m$, the Ricci tensor can be calculated as
\bea
R_{00} & = & - e^{-2A-2\a}\left[(m+1)(\ddot{D}+ \dot{D}^{2} -\dot{D}\dot{A}) + \sum_{k} d_k(\ddot{C_k}+ \dot{C_k}^{2} - \dot{C_k}\dot{A}) \right] \\
&+& e^{-2D-2\theta} \left[\a'' + \a'^2 - \a'\theta' + m\left(\th'+\frac{1}{r}\right)\a' + \sum_k d_k\b_k'\a'\right]\nonumber\\
R_{rr} &=& e^{-2A-2\a}\left[\ddot{D}- \dot{D}\dot{A} +(m+1)\dot{D}^2 + \sum_kd_k\dot{C}_k\dot{D} \right] \\
&-& e^{-2D-2\th}\left[\a'' + \a'^2 - \a'\theta' + m\left(\th''+\frac{\th'}{r}\right) + \sum_kd_k(\b_k''-\b_k'\th'+\b_k'^2)\right] \nonumber 
\eea
\bea
R_{x_iz_i} & = &
e^{-2A-2\a}\left[\ddot{C}_{i}+d_{i}\dot{C}^{2}_{i}-\dot{C}_{i}\dot{A}\,\,+\,(m+1)\dot{D}\dot{C}_{i}+\,
\sum_{k\not=i}d_{k}\dot{C}_{k} \dot{C}_{i}\right]\delta_{x_iz_i} \\
 &-& e^{-2D-2\theta} \left[\b_i'' +\b_i'\a'+d_i\b_i'^2-\b_i'\th' + m\left(\th' + \frac{1}{r}\right)\b_i' + 
\sum_{k\not=i}d_{k}\b_k'\b_i'\right]\delta_{x_iz_i}\nonumber \\
R_{\psi_1\psi_2} & = & e^{-2A-2\a}\left[ \ddot{D}\,+\,(m+1)\dot{D}^{2}-\dot{D}\dot{A}\,+\,\sum_{k}d_k\dot{C}_{k}\dot{D}\right]\delta_{\psi_1\psi_2} \\
&-&e^{-2D-2\theta} \left[\th'' +(2m-1)\frac{\th'}{r} + (m-1)\th'^2 + \left(\th'+\frac{1}{r}\right)\a'
+ \left(\th'+\frac{1}{r}\right)\sum_kd_k\b_k'\right]\delta_{\psi_1\psi_2} \, ,
\nonumber
\eea
where prime and dot denote differentiation with respect to $r$ and $t$ respectively. In addition to the above we also get a non-diagonal Ricci tensor which is
\be 
R_{0r} = e^{-A-D-\a-\th}\left[m\dot{D}\a' + \sum_k d_k\left(\dot{C}_k\a' + [\dot{D} -\dot{C}_k]\b_k'\right)\right]
\la{mix}
\ee

One can
simplify the above expressions by fixing $t$ and $r$ reparametrization invariances so
that
\bea
A &=& \sum_{k}d_{k}C_{k}\,+\,(m+1)D \,\, ,\\
\a &=& -\sum_kd_k\b_k - (m-1)\th \,\, .
\eea
With these gauge choices we have $\sqrt{-g}=e^{2A+2\th}r^m$ and  the diagonal components of the Ricci tensor become
\bea
R_{00} & = & e^{-2A-2\a}\left[-\ddot{A}+\dot{A}^{2}-(m+1)\dot{D}^{2}-\sum_{k}d_{k}\dot{C}_{k}^{2}\right]
+ e^{-2D-2\theta} \left[\a'' + \frac{m\a'}{r}\right]\\
R_{rr} &=& e^{-2A-2\a}\left[\ddot{D}\right] - e^{-2D-2\theta} \left[\th'' + \frac{m\th'}{r} + (m-1)\th'^2 + \a'^2 + \sum_kd_k\b_k'^2\right] \\
R_{x_iz_i} & = & e^{-2A-2\a}
\left[\ddot{C}_{i}\right]\delta_{x_iz_i} - e^{-2D-2\theta} \left[\b_i'' + \frac{m\b_i'}{r}\right]\delta_{x_iz_i}
\,\,\,\,\, i=1,2... \\
R_{\psi_1\psi_2} & = & e^{-2A-2\a}\left[\ddot{D}\right]\delta_{\psi_1\psi_2} - e^{-2D-2\theta} \left[\th'' + 
\frac{m\th'}{r}\right]\delta_{\psi_1\psi_2}
\eea

We assume that there are two $Dp$-branes and a single $Sq$-brane. Let us use the subscript $X=1,2$ to indicate the $Dp$-branes and the subscript $q$ for the $Sq$-brane. We 
assume separation of variables for the gauge potentials and the scalar field 
\bea
\la{sep1}
E_X &=& E_X^t(t) E_X^r(r)  \,\,\,\, , \,\,\,\, X=1,2 \,\, , \\
\la{sep2}
E_q &=& E_q^t(t) E_q^r(r) \,\, , \\
\la{sep3}
\f &=& \f_q(t) + \sum_{X=1}^2\f_X(r) \,\, .
\eea
From (\ref{eqn2}) we deduce
\bea
\label{e1}E_X^t &=& (V_X^t)^2 e^{2D-\e_Xa_X\f_q} \,\,\,\,\,\,\,\,\,\,\,\,\, , \,\,  X=1,2 \\
(E_X^r)' &=& \frac{(V_X^r)^2 e^{-\e_Xa_X\f_X} Q_X}{r^m}  \,\,\,\,\,\,\,\, , \,\, X=1,2 \\
\label{e2}E_q^r &=& \begin{cases} 
r^m(V_q^r)^2e^{2\a -2\th -\e_qa_q\sum_X\f_X} \,\,\, \hspace{1cm} \textrm{(Class I)} \cr 
1 \,\,\,\,\, \hspace{5cm}  \textrm{(Class II)}
\end{cases} \\
\dot{E}_q^t &=&  (V_q^t)^2 e^{-\e_qa_q\f_q} Q_q \hspace{2.5cm} \textrm{(Class I and II)} 
\eea
where $Q_X$'s and $Q_q$ are charges and we defined
\bea
V_X^r &=& e^{\a} e^{\sum_i \b_i} \, , \hspace{1cm}
V_X^t = e^{\sum_i C_i}  \hspace{1cm} \textrm{(for $i$ parallel to $p_X$)} \,\, , \,\, X=1,2\\
V_q^t &=& \begin{cases} 
e^{(m+1)D}e^{\sum_i C_i}  \hspace{1cm} \,\,\,\,\, \textrm{(Class I)} \cr
e^{\sum_i C_i} \,\, \hspace{2.5cm} \textrm{(Class II)} 
\end{cases} \hspace{1.3cm} \textrm{(for $i$ parallel to $q$)} \\
V_q^r &=& 
\begin{cases} 
e^{(m+1)\th}e^{\sum_i \b_i} \hspace{1.5cm} \textrm{(Class I)} \cr
e^{\sum_i \b_i} \,\,\, \hspace{2.5cm} \textrm{(Class II)} 
\end{cases}  \hspace{1.2cm} \textrm{(for $i$ parallel to $q$)} 
\eea
In order to separate time and radial parts completely in the Einstein's equations (\ref{eqn1}) with the above ansatz (\ref{sep1})-(\ref{sep3}), we need to impose
\bea
\la{e3}e^{\e_Xa_X\f_q} &=& (V_X^t)^2 e^{4D-2A} \,\,\,\,\,\,\,\, , \,\,  X=1,2 \\
\la{e4}e^{\e_qa_q\sum_X\f_X} &=& 
\begin{cases} 
(V_q^r)^2 e^{4\a -4 \theta} \hspace{2cm}   \textrm{(Class I)} \cr
(V_q^r)^2 \hspace{3cm}  \, \textrm{(Class II)} 
\end{cases}
\eea
Putting these restrictions in (\ref{e1}) and (\ref{e2}) we find
\bea
\la{fin1}
E_X^t &=& e^{2A-2D} \,\,\,\,\,\,\, , \,\,\,\, X=1,2 \\
\la{fin2}
E_q^r &=& \begin{cases} 
r^m e^{2\th -2\a} \, \hspace{1.1cm} \textrm{(Class I)} \cr
1 \hspace{2.5cm} \textrm{(Class II)} 
\end{cases}
\eea
Let us now define
\bea
\d_X^i &=& \begin{cases} (d-p_X-3) \hspace{1cm} \textrm{(for $i$ parallel to $p_X$)} \,\,\,\, , \,\, X=1,2 \cr -(p_X +1) \hspace{1.4cm} \textrm{(otherwise)} \,\,\,\,\,\,\,\, , \hspace{1.4cm} X=1,2
\end{cases} \\
\d_q^i &=& \begin{cases} (d-q-3) \hspace{1.3cm} \textrm{(for $i$ parallel to $q$)} \cr -(q +1) \hspace{1.7cm} \textrm{(otherwise)} \end{cases}
\eea

After these we obtain the following set of equations from (\ref{eqn1}) involving radial derivatives:
\bea
\a '' + \frac{m\a'}{r} &=& \frac{1}{2}\sum_{X=1}^2 \frac{(d-p_X-3)e^{-\e_Xa_X\f_A}(V_X^r)^2 Q_X^2}{(d-2)r^{2m}}  \\
\b_i'' + \frac{m\b_i'}{r} &=& \frac{1}{2}\sum_{X=1}^2 \frac{\d_X^i e^{-\e_Xa_X\f_X}(V_X^r)^2 Q_X^2}{(d-2)r^{2m}}  \\
\th'' +\frac{m\th'}{r} &=& -\frac{1}{2}\sum_{X=1}^2 \frac{(p_X+1)e^{-\e_Xa_X\f_X}(V_X^r)^2 Q_X^2}{(d-2)r^{2m}}  
\eea
\bea
\th'' + \frac{m\th'}{r} &=& - (m-1)\th'^2 - \a'^2 - \sum_k d_k\b_k'^2  -\frac{1}{2}
\sum_{X=1}^{2} [(\f_X)']^2 \nonumber \\
&+& \frac{1}{2}\sum_{X=1}^2\frac{(d-p_X-3)e^{-\e_Xa_X\f_X}(V_X^r)^2 Q_X^2}{(d-2)r^{2m}} \\
(E_X^r)' &=& \frac{(V_X^r)^2 e^{-\e_Xa_X\f_X} Q_X}{r^m}  \,\,\,\, , \hspace{2.3cm} X=1,2  \\
(\f_X)'' + \frac{m (\f_X)'}{r} &=& -\frac{\e_Xa_X e^{-\e_Xa_X\f_X} (V_X^r)^2 Q_X^2}{2r^{2m}} \,\,\,\, , \, \hspace{1cm} 
X=1,2
\eea
where the last one is the dilaton equation (\ref{eqn2}).
These equations are exactly those that are obtained for an intersecting $p_1-p_2$ brane system whose solution is given in the main text in (\ref{solmet})-({\ref{pscalar}).

The set of equations containing time derivatives is as follows:
\bea
\ddot{C_i} &=& -\frac{\d_q^i (V_q^t)^2 e^{-\e_qa_q\f_q}Q_q^2}{2(d-2)} \\
\ddot{D} &=& \frac{(q+1) (V_q^t)^2 e^{-\e_qa_q\f_q}Q_q^2}{2(d-2)}  \\
\ddot{A} &=& \dot{A}^2 - (m+1)\dot{D}^2 -\frac{(\dot{\f}_q)^2}{2} - \frac{(d-q-3) (V_q^t)^2 e^{-\e_qa_q\f_q}Q_q^2}{2(d-2)} - \sum_kd_k\dot{C}_k^2  \\
\dot{E}_q^t &=&  (V_q^t)^2 e^{-\e_qa_q\f_q} Q_q  \\
\ddot{\f}_q &=& \frac{\e_qa_q(V_q^t)^2 e^{-\e_qa_q\f_q}Q_q^2}{2} \,\, ,
\eea
where the last one is from the dilaton (\ref{eqn2}) field equation. These are nothing but the equations of motion for a single $Sq$-brane whose solution is given in (\ref{smetric})-(\ref{sson}).

The only thing left is to analyze the separability conditions (\ref{e3})-(\ref{e4}) and the $R_{0r}$ component of the Ricci tensor (\ref{mix}). The separability requirements fix the dimension of the smeared directions of each $Dp$-brane that are intersecting (Class II) or non-intersecting (Class I) with the $Sq$-brane which are given in (\ref{double2}), (\ref{double1}), (\ref{triple21}), (\ref{triple22}), (\ref{triple11}) and (\ref{triple12}). After some algebra, it can be shown that even though one has 3 equations  only two of them are independent. This is actually not surprising since in a triple intersection there are 7 unknowns in the metric (\ref{metric2}) and given $p_1$, $p_2$, $q$, $d$ and the intersection dimension of $p_1$ and $p_2$, only two more relations are allowed.
Moreover, with these Einstein's equation for the $R_{0r}$ component (\ref{mix}) is solved automatically with an additional restriction on integration constants which are given in (\ref{constant2}), (\ref{constant}),(\ref{constant4}) and  (\ref{constant3}). This restriction is also necessary for (\ref{e3}).
The final forms of the gauge potentials are obtained from (\ref{radial}), (\ref{ssonson}), (\ref{fin1}) and (\ref{fin2}) as
\bea
\la{gauge1}
E_{X}(r,t) &=& e^{2A(t)-2D(t)} H_{X}^{-1}(r) \,\,\,\, , \,\,\,\, X=1,2\\
\la{gauge2}
E_q(r,t) &=& \begin{cases} 
\frac{c_3 e^{c_3(t-t_0)}}{Q_q\cosh[c_3(t-t_0)]} r^mH_1(r)H_2(r) \,\,\,\,\,\,\,\, \textrm{(Class I)} \cr
\frac{c_3 e^{c_3(t-t_0)}}{Q_q\cosh[c_3(t-t_0)]} \,\,\,\,\,\, \hspace{2.5cm} \textrm{(Class II)} 
\end{cases}
\eea

The whole analysis of this section is still valid when there is only a single $Dp$-brane. 
Simply dropping the subindex $X$ and setting one of the $H_X$'s to 1 are enough.

\end{document}